\documentclass[oldversion]{aa}

\usepackage{graphicx}
\begin{document}
\title{The candidate filament close to the 3C295 galaxy cluster: optical
and X-ray spectroscopy \thanks{Based on observations collected with
the Telescopio Nazionale Galileo, Roque de Los Muchachos, La Palma, TF
- Spain. and with the {\it Chandra} X-ray observatory}}

\author{Valerio D'Elia$^1$, Fabrizio Fiore$^1$, Smita Mathur$^2$
Filomena Cocchia$^1$
}
\institute {INAF-Osservatorio Astronomico di Roma \\
via Frascati 33, Monteporzio-Catone (RM), I00040 Italy,
\and
The Ohio State University \\ 
Columbus, OH 43210, USA.}
\date{November, 28 2007}

\abstract {We present a detailed analysis of the overdensity of X-ray
sources colse to the 3C 295 galaxy cluster ($z=0.46$) to assess
whether it is associated with a filament of the large-scale structure
of the Universe. We obtained optical spectra of the optical
counterparts of eleven sources associated with the filament, finding
that one is at $z=0.474$. This is a type 1 AGN at $1.5$ arcmin from
the cluster center.  We found three more sources with a redshift in
the range $0.37 \div 0.53$. We extracted the stacked X-ray spectrum of
{\bf 47} X-ray sources belonging to the putative filament. We found a
significant narrow (at the resolution of the {\it Chandra} ACIS-I)
line at E$\sim4.4$keV, the energy of the iron K$\alpha$ line at the
redshift of the cluster. The detection of this line is confirmed at a
confidence level of better than $3\sigma$ and its energy is
constrained to be in the range 6.2--6.47 (at a 90\% confidence level),
excluding an identification with the 6.7 helium-like iron line from
the hot cluster ICM at better than 4 $\sigma$. We conclude that the
detection of the redshifted line is a strong indication that at least
several of the excess sources lie at z$\sim0.46$ and that AGNs are
efficient tracers of the ``filament'' connected with the central
cluster of galaxies.


\authorrunning {D'Elia et al.}
\titlerunning {A Candidate Filament Near 3C 295}
}
\maketitle

\section{Introduction}

N-body and hydrodynamical simulations show that clusters of galaxies
lie at the nexus of several filaments of galaxies (see e.g. Peacock
1999, Dav\`e et al. 2001 and references therein).  Such a ``cosmic web''
of voids and filaments maps out the large-scale structure (LSS) of the
Universe. According to the same simulations, these filaments contain a
large fraction ($30-40\%$) of the baryons in the Universe at z$<1$;
the remainder end up in the hot gas in clusters of galaxies, and
in stars and cold gas clouds.  Despite its larger total mass,
observations of the intergalactic matter in filaments have
yielded only limited information, mostly due to its low density: most
of the baryons in this phase should be at densities only 10-100 times
higher than the average density of the Universe. The most direct way
to detect a filament at low redshift is by  means of its soft X-ray diffuse
emission (see e.g. Zappacosta et al. 2002, Soltan, Freyberg \&
Hasinger 2002), or using soft X-ray and UV absorption-line studies
(see e.g. Fiore et al. 2000 and references therein, Nicastro et
al. 2002, Nicastro et al. 2003, Mathur et al. 2003, Nicastro 2005,
Nicastro 2005b).  Both methods require very difficult observations, at
the limit of the present generation of X-ray and UV facilities.

Alternatively, filaments could be mapped out by galaxies (Daddi et
al. 2001, Giavalisco \& Dickinson 2001) and by the much more luminous
Active Galactic Nuclei (AGNs), assuming that AGNs trace
galaxies. Since rich clusters of galaxies are good indicators of
regions of sky where filaments converge, numerous AGN searches around
clusters of galaxies have been performed in the past.  Several of
these studies suggest that overdensities of AGNs exist around distant
clusters of galaxies (Molnar et al. 2002 for the cluster Abell 1995;
Best et al. 2002 for MS1054-03; Martini et al. 2002 for Abell 2104;
Pentericci et al. 2002 for the protocluster at $z \sim 2.16$ around
the radio galaxy MRC 1138-206; see also Almaini et al. 2003 for the
ELAIS North field). Cappelluti et al. (2005) performed the first
systematic study of serendipitous X-ray source density around $10$
high-z clusters; they found a factor of approximately two overdensity
in $4$ cluster fields.  Many of these studies have been performed in
X-rays, since extragalactic X-ray sources, which are mostly AGNs, have
a space density $\sim 10$ times higher than optically-selected AGNs
(see Yang et al. 2003), and therefore provide denser tracing of
LSS. Later studies suggested that AGNs are more frequently found close
to rich clusters than previously thought.  Ruderman and Ebeling (2005)
studied the spatial distribution of a $95 \% $ complete sample of
$508$ X-ray point sources detected in the $0.5 - 2$ KeV band by {\it
Chandra} ACIS-I close to $51$ massive galaxy clusters ($z \sim 0.3
\div 0.7$). They found a significant ($8 \sigma$) excess within $3.5$
Mpc of the cluster center, that they attribute to AGNs connected to
the central clusters. Similarly, Martini et al. (2006)
spectroscopically-identified $40$ sources in the field of $8$ {\it
Chandra} low-redshift clusters ($0.05 < z < 0.31$). Using a
combination of optical emission lines, X-ray spectral shapes, and
multiwavelength flux ratios, they concluded that at least $35$ of
these objects have AGN signatures.  Dilution of low-luminosity AGN
spectral signatures by host-galaxy starlight and obscuration of
accretion onto the central black hole do not easily allow sources to
be optically classified as AGNs. Even though they refer to cluster
members, we stress that the spectroscopic identification of sources
close to clusters cannot be an easy task.

In this paper, we study the X-ray source excess close to the 3C 295
galaxy cluster.  Cappi et al.  (2001) observed the {\it Chandra}
$8\times 8$ arcmin ACIS field around 3C 295 ($z=0.46$).  They reported
the tentative $0.5-2$ keV detection of an overdensity of faint X-ray
sources in a region of a few arcmin around the cluster, with respect
to the average X-ray source density at the same flux limit.  However,
the observation was too short ($\sim 18$ ks), and the source samples
consequently too small, to derive any more detailed conclusion.
D'Elia et al. (2005) observed the 3C 295 field using {\it Chandra} for
about 100ks.  They performed a detailed study of the field, applying
three different methods of analysis, namely, the logN-logS computed
separately for each of the four ACIS-I chips, the two point,
two-dimensional Kolmogorov-Smirnov (KS) test, and the angular
correlation function of the field. All of these analysis methods
suggest an asymmetric distribution of sources in the NE region of the
field and a strong clustering on scales of a few arcmins.

The aim of this work is to investigate whether the asymmetric
distribution of X-ray sources close to the 3C 295 galaxy cluster, found
by D'Elia et al. (2005), is associated with a ``filament''  of the LSS of
the Universe and connected to the central cluster. The paper is
organized as follows: Section 2 presents the optical identifications
of the X-ray {\it Chandra} sources; Section 3 presents the analysis of
the X-ray stacked spectrum of the sources in the overdensity region;
Section 4 discusses the results and draws our conclusions.  We adopt
in the following a concordance cosmology with $H_0 =65$ km/s Mpc,
$\Omega_M = 0.3$ and $\Omega_{\Lambda} = 0.7$, (Spergel et al. 2003).

\begin{figure}
\centering
\includegraphics[angle=0,width=11cm]{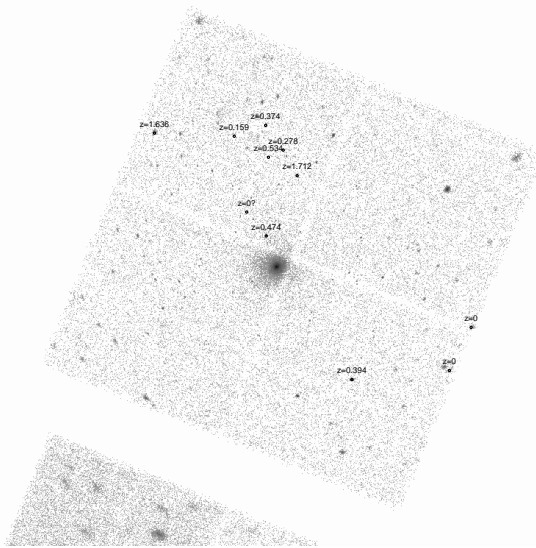}
\caption{The position of the spectroscopically-identified X-ray
sources inside the 3C 295 field.}
\label{spe1}
\end{figure}

\begin{figure}
\centering
\includegraphics[angle=-90,width=8cm]{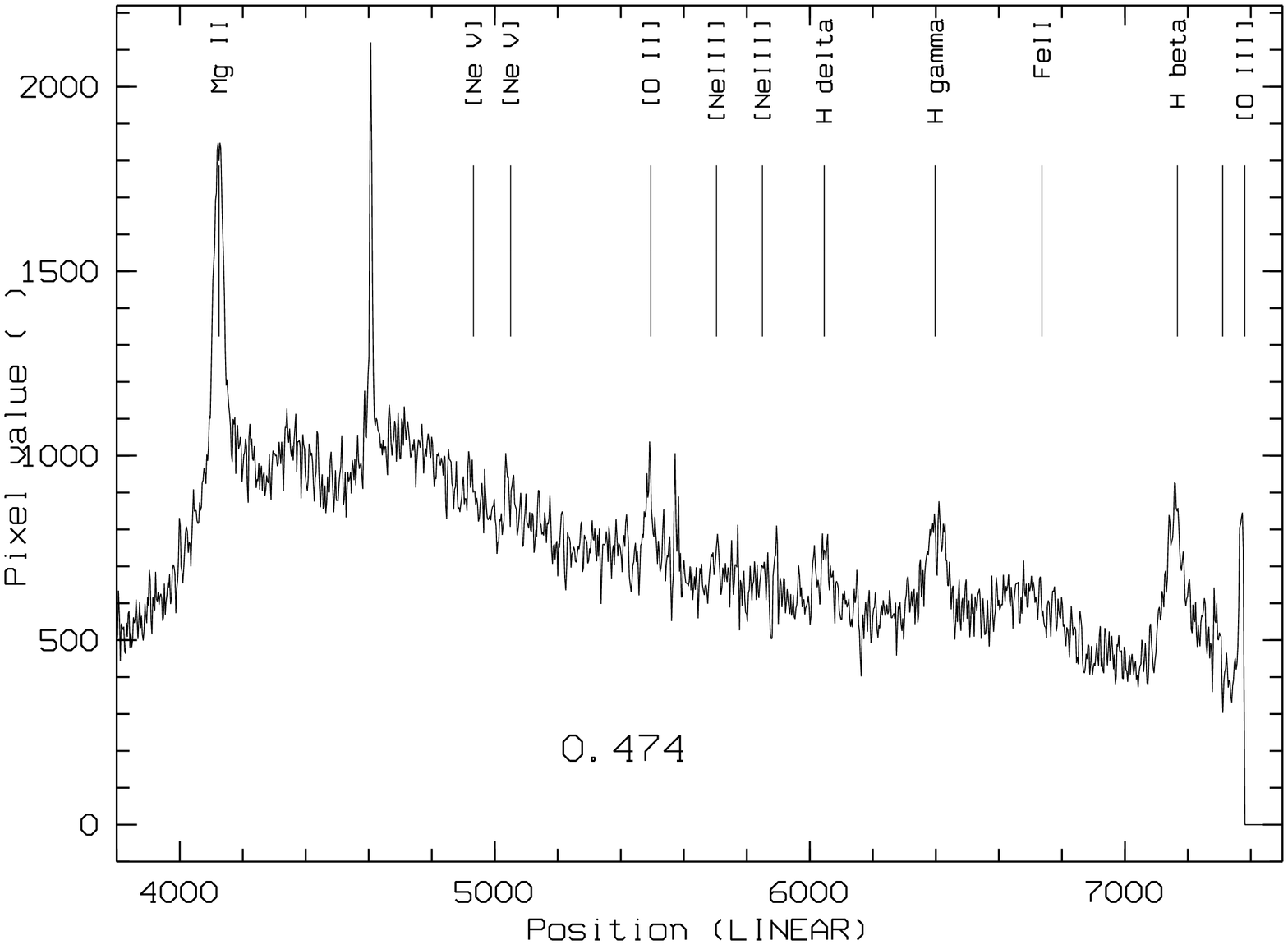}
\includegraphics[angle=-90,width=7.5cm]{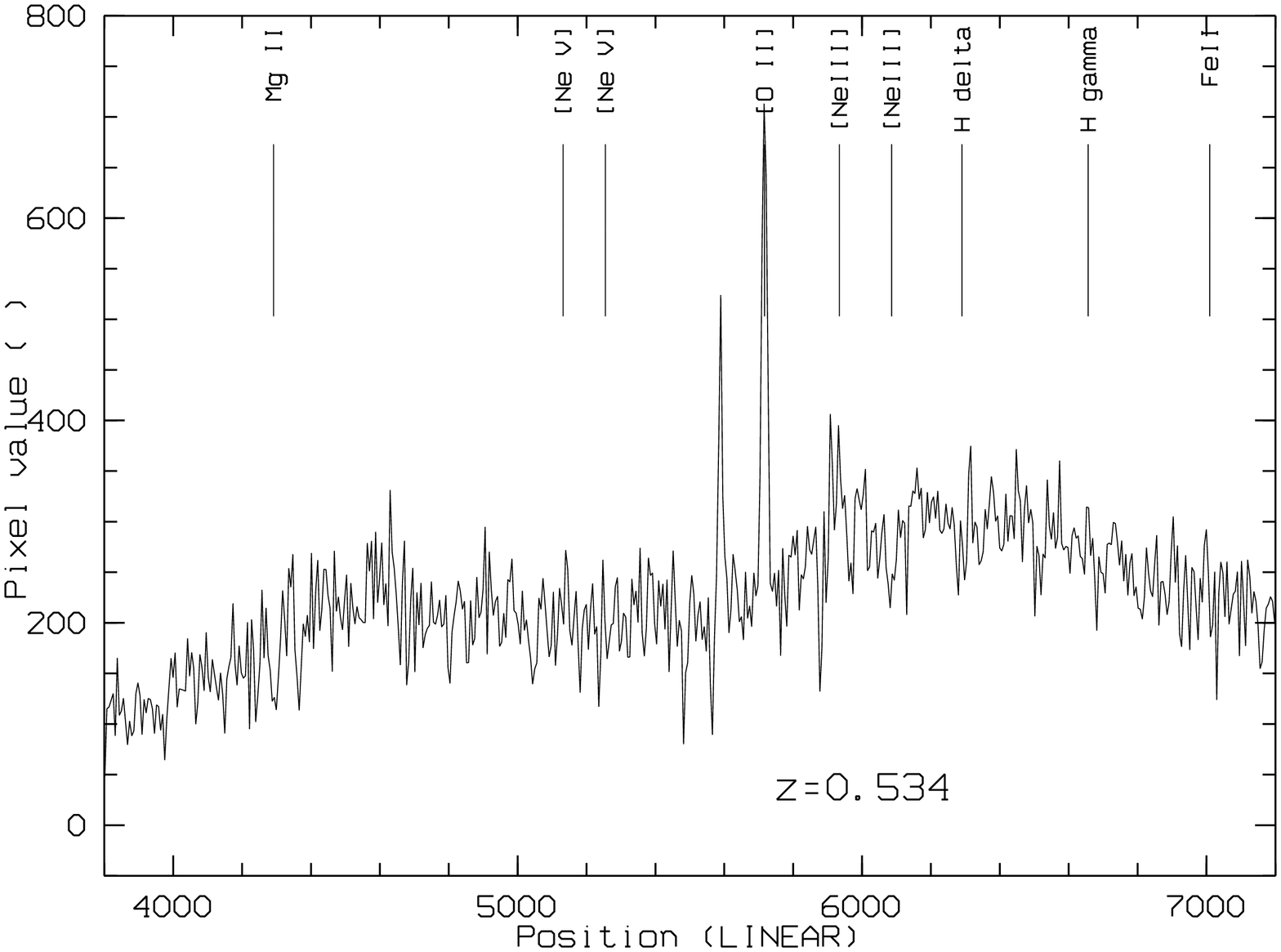}
\includegraphics[angle=-90,width=8cm]{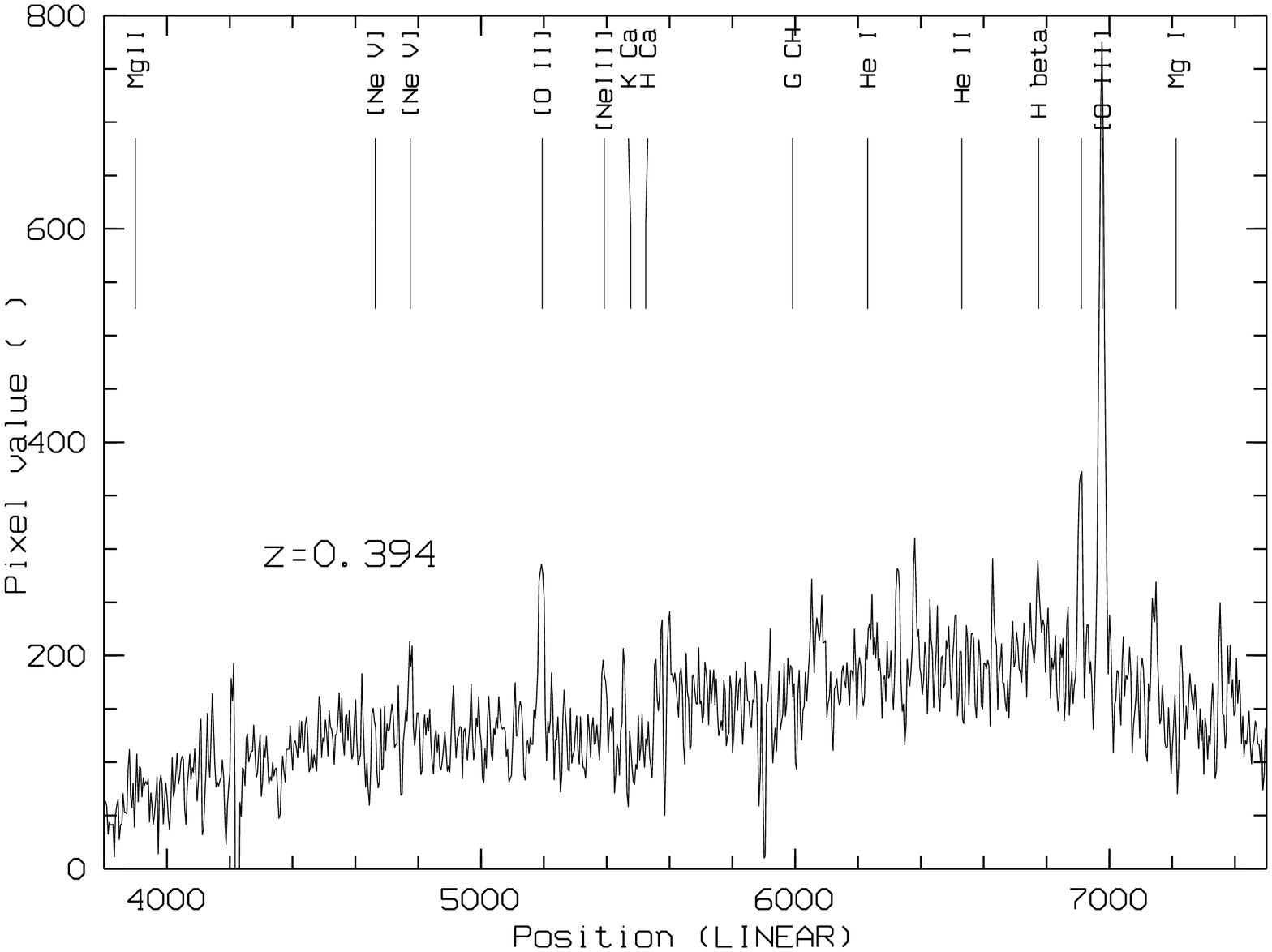}
\includegraphics[angle=-90,width=8cm]{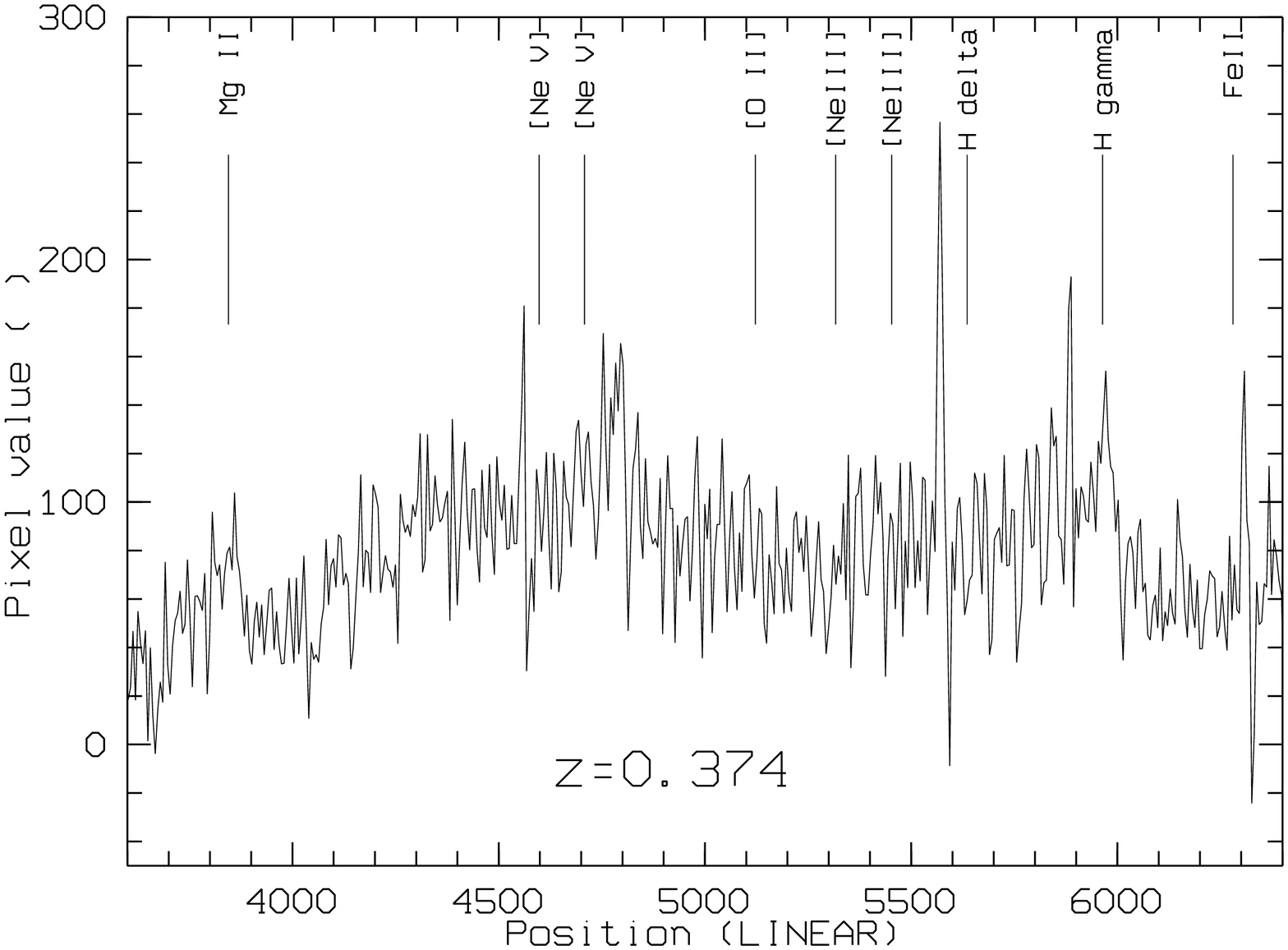}
\caption{The optical spectra of the $4$ sources whose redshift is closer
to that of the central cluster of the 3C 295 field ($z = 0.46$). }
\label{spe1}
\end{figure}

\section{Optical Analysis}

The asymmetric distribution of the sources around the 3C 295 galaxy
cluster and the overdensity in the NE region of the 3C 295 field found
by D'Elia et al. (2005) posed the obvious question of whether such
overdensities are due to background or foreground objects, or if the
excess sources are spatially linked to the central cluster along a
filmament of the LSS of the Universe. The natural way to answer this
question is to optically identify the X-ray sources and determine
their redshift by means of an analysis of the spectrum.

\subsection{Data Acquisition and Reduction}

To acquire the optical spectra of the 3C 295 sources, we proposed a
Multi Object Spectroscopy (MOS) programme using DOLORES@TNG (AOT 9).
The project was accepted and two nights of observation were
allocated. Two acquisition images each of exposure time $900$ s and
with a limiting magnitude of $R \sim 24$, had already been obtained
using DOLORES on March 19, 2002. These images were centered on the NE
and SW quadrants of the {\it Chandra} field, and covered an area of
9'X9', with respect to the 16'X16' area of the four ACIS-I chips.  We
acquired the spectral data on 22 and 23 March, 2004.  We positioned in
slit $26$ optically-identified X-ray sources whose optical counterpart
was brighter than $R=22$. Unfortunately, due to weather and technical
difficulties, we were able to observe our target field for only six
hours.

All spectra acquired were collected into three MOS images.  The data
were reduced using the IRAF software and its standard analysis
packages. Each MOS image was corrected for the electronic bias; the
spectra were then separated, flatfielded and
background-subtracted. Finally, we removed cosmic rays and performed a
wavelength calibration for each spectrum.

\subsection{Spectral Analysis and Redshift Determination}

Table 1 summarizes the results of our optical identification campaign.
In detail, we show the X-ray and optical coordinates, the R
magnitude, and the $0.5-10$ keV X-ray flux. The identification number
given in the first column refers to the whole $0.5-10$ keV sample in
D'Elia et al. 2005.  Column $8$ indicates the {\it Chandra} ACIS-I
chip to which each X-ray source belongs, while column $9$ indicates a
source with its optical counterpart, selected for the stacked X-ray
spectrum, as described in the next section.

We were able to obtain reliable redshifts for only $11$ of the sources
placed inside the slit.  This is because the optical counterparts of
several X-ray objects were too faint for a reasonable signal-to-noise
ratio to be achieved in the observing time the seeing conditions.  The
last three columns of Table 1 report the redshift, the X-ray ($0.5
-10$ keV) luminosity, and the classification of the
spectroscopically-identified sources.

Fig. 1 shows the position of these sources inside the 3C 295
field. Three of the $11$ sources were found to be stars, and $2$ of
these are located in the SW part of the 3C 295 field. This enhances
the asymmetric distribution found by D'Elia et al. (2005), who showed
that the SW region was underpopulated, with only $12$ X-ray sources
identified in the $0.5 - 10$ keV band of a total of $121$ sources in
the entire field. The star identified in the NE chip cannot be the
true optical counterpart of the X-ray source, because a
fainter source, inaccessible to 4m telescope spectroscopy, lies
within the X-ray error box; a question mark identifies this source in
Fig. 1.

\begin{figure}
\centering
\includegraphics[angle=0,width=11cm]{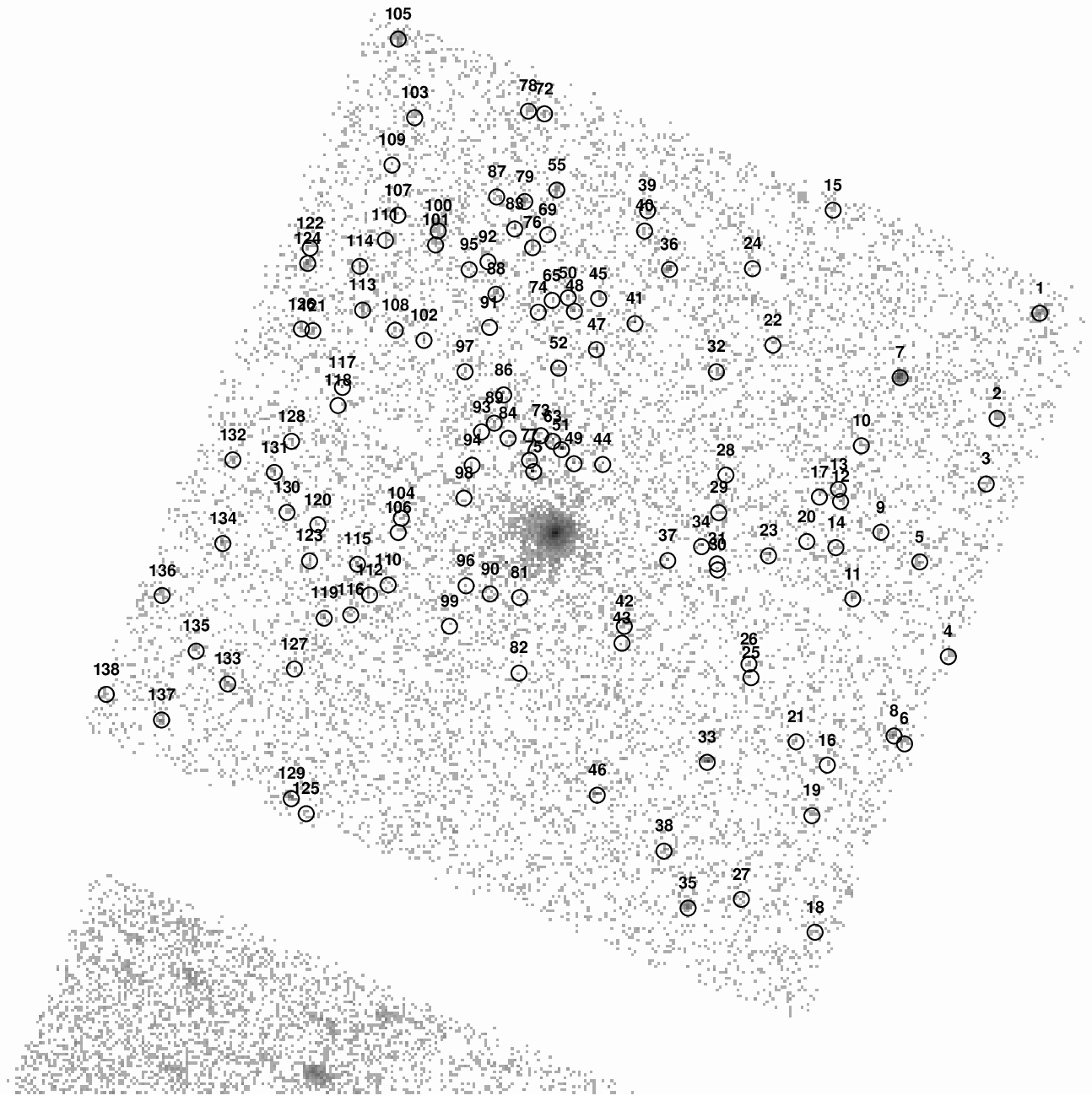}
\includegraphics[angle=0,width=11cm]{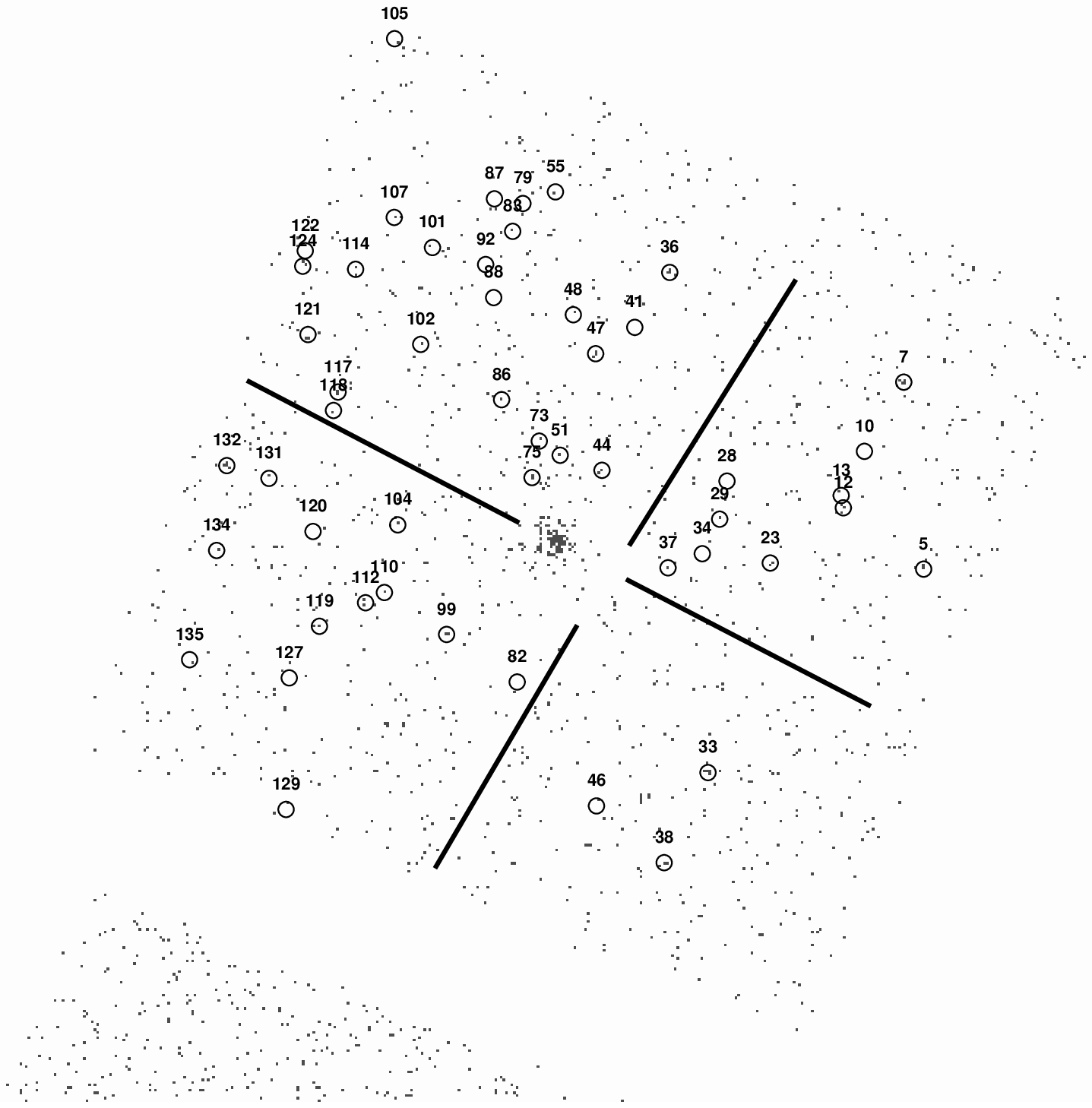}
\caption{Top panel: the $0.5 - 10$ keV {\it Chandra} image of the 3C 295
field; circles represent the $121$ sources identified with the PWDetect
algorithm. Bottom panel: the $4.1 - 4.9$ keV image of the 3C 295 field;
circles represent the $51$ sources selected for the production of the stacked
X-ray spectrum, and black lines divide the image in its four quadrants.}
\label{spe1}
\end{figure}

One of the sources in the NE region has a redshift of $0.474$ (see
Fig. 1 upper panel). This is a type 1 AGN at $1.5$ arcmin from the center
of the cluster, implying a physical distance of $600$ kpc. The spread in
velocity with respect to the cluster redshift is $4200$ km/s.  Three more
sources have a redshift in the range $0.37 \div 0.53$: two of
these are located in the NE region, while one is in the SW
region. Their spectra are shown in Fig. 2. The remaining sources are
two $z>1$ QSOs and two lower redshift galaxies ($z <0.3$).  

The identification of only 1 source at the redshift of 3C295 is
unsurprising.  D'Elia et al. (2005) showed that the 0.5-10 keV
luminosity of the sources in the 3C295 field should be in the range
$7.5\times 10^{41} \div 10^{44}$ ergs s$^{-1}$ if they were associated
to filaments connected to the central cluster of galaxies. Therefore
several starburst galaxies and low luminosity AGNs are expected in
this structure. Assuming a typical X-ray to optical flux ratio
(e.g. Fiore et al. 2003), one would expect that their optical
magnitude would be fainter than R=22. These magnitudes were beyond the
reach of our 4m-class telescopes. Our spectroscopic run was able to
identify only optically unobscured QSOs and bright Seyfert galaxies at
z$\sim0.46$. At similar redshift, the identification of sources that
are both optically fainter and highly-obscured will require data from
8m-class telescopes.

Fortunately, optical spectroscopy is not the only tool available to
search for evidence of the association of the X-ray sources with the
large-scale structure linked to 3C295. In the next section, we
perform a new test based on X-ray spectroscopy.


\begin{table*}[ht]
\caption{\bf Optical identification of the x-ray sources in the 3c295 field }
\footnotesize
\smallskip
\begin{tabular}{|l|ccccccccccc|}
\hline
\multicolumn{1}{|l|}{}&
\multicolumn{2}{|c|}{X position}&
\multicolumn{2}{|c|}{Optical position}&
\multicolumn{1}{|c|}{}&
\multicolumn{1}{|c|}{X flux (cgs)}& 
\multicolumn{1}{c}{}&
\multicolumn{1}{c}{}&
\multicolumn{1}{c}{}&
\multicolumn{1}{c}{}&
\multicolumn{1}{c|}{} \\
\hline
\multicolumn{1}{|l|}{n}&
\multicolumn{1}{|c|}{RA}&
\multicolumn{1}{|c|}{DEC}&
\multicolumn{1}{|c|}{RA}&
\multicolumn{1}{|c|}{DEC}&
\multicolumn{1}{|c|}{RMag}&
\multicolumn{1}{|c|}{0.5-10 keV}&
\multicolumn{1}{|c|}{Chip}&
\multicolumn{1}{|c|}{Stack}&
\multicolumn{1}{|c|}{z}&
\multicolumn{1}{|c|}{LX (erg/s)}&
\multicolumn{1}{|c|}{Class}\\

\hline

   2   &     14:10:17.329 & 52:14:39.7  &  14:10:17.49 & 52:14:38.55 & 24.43 &  1.60E-15  &  2   &    n   &  	-     &  - &            \\
   3   &     14:10:18.901 & 52:13:13.98 &  14:10:19.17 & 52:13:14.99 & 18.77 &  1.90E-15  &  2   &    n   &  	-     &  - &            \\
   4   &     14:10:24.356 & 52:9: 28.65 &  14:10:24.52 & 52:9:27.53  & 16.6  &  3.98E-15  &  2   &    n   &  	0     &  - &    Star    \\
   5   &     14:10:28.38  & 52:11:32.56 &  14:10:28.58 & 52:11:33.25 & 21.06 &  2.34E-15  &  2   &    y   &  	-     &  - &            \\
   6   &     14:10:30.608 & 52:7:34.86  &  14:10:30.91 & 52:7:35.8   & 17.77 &  2.94E-15  &  3   &    n   &  	0     &  - &	Star    \\
   7   &     14:10:31.096 & 52:15:33.3  &  14:10:31.33 & 52:15:33.9  & 20.32 &  4.54E-14  &  2   &    y   &  	-     &	 - &	        \\
   8   &     14:10:32.098 & 52:7:45.19  &  14:10:32.38 & 52:7:45.74  & 23.62 &  8.84E-15  &  3   &    n   &  	-     &	 - &	        \\
   9   &     14:10:33.891 & 52:12:11.35 &  14:10:34.15 & 52:12:11.6  & 23.6  &  2.59E-15  &  2   &    n   &  	-     &	 - &            \\
  11   &     14:10:37.911 & 52:10:44.4  &  14:10:38.15 & 52:10:44.87 & 21.91 &  4.21E-15  &  2   &    n   &  	-     &  - &	        \\
  13   &     14:10:39.906 & 52:13:7.61  &  14:10:39.91 & 52:13:8.99  & 24.2  &  1.02E-15  &  2   &    y   &  	-     &	 - &	        \\
  14   &     14:10:40.287 & 52:11:51.43 &  14:10:40.56 & 52:11:52.16 & 23.5  &  9.04E-16  &  2   &    n   &  	-     &	 - &	        \\
  16   &     14:10:41.578 & 52:7:7.35   &  14:10:41.59 & 52:7:5.62   & 23.55 &  8.60E-16  &  3   &    n   &  	-     &	 - &	        \\
  17   &     14:10:42.614 & 52:12:57.53 &  14:10:42.55 & 52:12:57.11 & 25.1  &  6.32E-16  &  2   &    n   &  	-     &	 - &	        \\
  20   &     14:10:44.417 & 52:11:59.43 &  14:10:44.63 & 52:12:0.07  & 21.94 &  5.42E-16  &  2   &    n   &  	-     &	 - &	        \\
  21   &     14:10:45.979 & 52:7:37.84  &  14:10:45.98 & 52:7:40.67  & 21.66 &  2.05E-15  &  3   &    n   &  	-     &	 - &	        \\
  23   &     14:10:49.855 & 52:11:40.85 &  14:10:49.63 & 52:11:38.53 & 23.35 &  5.98E-15  &  2   &    y   &  	-     &	 - &            \\
  25   &     14:10:52.372 & 52:9:1.76   &  14:10:52.68 & 52:9:1.61   & 23.91 &  6.46E-16  &  3   &    n   &  	-     &  - &	        \\
  28   &     14:10:55.898 & 52:13:26.4  &  14:10:56.23 & 52:13:26.15 & 23.89 &  4.34E-16  &  2   &    y   &  	-     &	 - &	        \\
  29   &     14:10:56.927 & 52:12:37.4  &  14:10:56.96 & 52:12:36.36 & 23.4  &  1.99E-15  &  2   &    y   &  	-     &	 - &	        \\
  30   &     14:10:57.079 & 52:11:22.48 &  14:10:57.27 & 52:11:23.7  & 24.5  &  9.13E-16  &  3   &    n   &  	-     &	 - &	        \\
  31   &     14:10:57.168 & 52:11:30.48 &  14:10:57.41 & 52:11:30.71 & 21.54 &  2.76E-15  &  2   &    n   &  	-     &	 - &            \\
  33   &     14:10:58.606 & 52:7:11.21  &  14:10:58.79 & 52:7:12.03  & 20.52 &  3.49E-14  &  3   &    n   &  	0.394 &  1.88E43 &	QSO     \\
  34   &     14:10:59.369 & 52:11:52.94 &  14:10:59.61 & 52:11:53.37 & 22.27 &  4.84E-15  &  2   &    y   &  	-     &	 - &	        \\
  36   &     14:11:3.894  & 52:17:55.07 &  14:11:4.19  & 52:17:55.1  & 19.12 &  1.45E-14  &  1   &    y   &  	-     &	 - &	        \\
  37   &     14:11:4.172  & 52:11:34.83 &  14:11:4.35  & 52:11:35.38 & 22.37 &  1.41E-15  &  3   &    y   &  	-     &	 - &	        \\
  39   &     14:11:7.011  & 52:19:11.86 &  14:11:7.3   & 52:19:12:   & 22.93 &  1.64E-15  &  1   &    n   &  	-     &	 - &	        \\
  41   &     14:11:8.803  & 52:16:44.4  &  14:11:9.1   & 52:16:44.8  & 22.8  &  6.03E-15  &  1   &    y   &  	-     &	 - &	        \\
  43   &     14:11:10.674 & 52:9:46.88  &  14:11:10.56 & 52:9:45.38  & 24.5  &  1.10E-15  &  3   &    n   &  	-     &	 - &	        \\
  44   &     14:11:13.41  & 52:13:40.4  &  14:11:13.67 & 52:13:40.96 & 20.92 &  4.60E-15  &  1   &    y   &  	-     &	 - &	        \\
  46   &     14:11:14.193 & 52:6:28.41  &  14:11:14.41 & 52:6:28.84  & 20.61 &  1.19E-14  &  3   &    y   &  	-     &	 - &            \\
  47   &     14:11:14.296 & 52:16:10.49 &  14:11:14.61 & 52:16:10.82 & 20.93 &  6.76E-15  &  1   &    n   &  	1.712 &  1.36E44 &	QSO     \\
  48   &     14:11:17.424 & 52:17:0.67  &  14:11:17.7  & 52:17:2.11  & 22.24 &  1.99E-15  &  1   &    y   &  	-     &	 - &            \\
  50   &     14:11:18.303 & 52:17:18.17 &  14:11:18.8  & 52:17:15.81 & 19.79 &  5.40E-16  &  1   &    n   &  	0.278 &  1.61E42 &	Gal     \\
  51   &     14:11:19.271 & 52:13:59.77 &  14:11:19.59 & 52:14:0.37  & 20.81 &  1.79E-14  &  1   &    y   &  	-     &	 - &	        \\
  52   &     14:11:19.669 & 52:15:46.07 &  14:11:20:00 & 52:15:46.6  & 23.0  &  1.78E-15  &  1   &    n   &  	-     &	 - &	        \\
  63   &     14:11:20.497 & 52:14:10.64 &  14:11:20.79 & 52:14:11.59 & 15.07 &  1.12E-15  &  1   &    n   &  	-     &	 - &	        \\
  65   &     14:11:20:6   & 52:17:14:75 &  14:11:20:86 & 52:17:15:97 & 19:13 &  7.52E-16  &  1   &    n   &  	-     &	 - &	        \\
  69   &     14:11:21.218 & 52:18:40.47 &  14:11:21.65 & 52:18:39.03 & 19.32 &  9.43E-16  &  1   &    n   &  	-     &	 - &	        \\
  73   &     14:11:22.199 & 52:14:17.99 &  14:11:22.5  & 52:14:18.6  & 23.9  &  4.02E-16  &  1   &    y   &  	-     &	 - &            \\
  74   &     14:11:22.574 & 52:16:59.34 &  14:11:22.85 & 52:17:0.44  & 20.27 &  4.69E-16  &  1   &    n   &  	0.534 &  2.23E41 &    Gal     \\

\hline
\end{tabular}

\normalsize
\end{table*}

\begin{table*}[ht]
\footnotesize
\smallskip
\begin{tabular}{|l|ccccccccccc|}
\hline
\multicolumn{1}{|l|}{}&
\multicolumn{2}{|c|}{X position}&
\multicolumn{2}{|c|}{Optical position}&
\multicolumn{1}{|c|}{}&
\multicolumn{1}{|c|}{X flux (cgs)}& 
\multicolumn{1}{c}{}&
\multicolumn{1}{c}{}&
\multicolumn{1}{c}{}&
\multicolumn{1}{c}{}&
\multicolumn{1}{c|}{} \\
\hline
\multicolumn{1}{|l|}{n}&
\multicolumn{1}{|c|}{RA}&
\multicolumn{1}{|c|}{DEC}&
\multicolumn{1}{|c|}{RA}&
\multicolumn{1}{|c|}{DEC}&
\multicolumn{1}{|c|}{RMag}&
\multicolumn{1}{|c|}{0.5-10 keV}&
\multicolumn{1}{|c|}{Chip}&
\multicolumn{1}{|c|}{Stack}&
\multicolumn{1}{|c|}{z}&
\multicolumn{1}{|c|}{LX (erg/s)}&
\multicolumn{1}{|c|}{Class}\\

\hline


  75   &     14:11:23.209 & 52:13:31.04 &  14:11:23.52 & 52:13:31.8  & 19.66 &  5.65E-14  &  1   &    n   &  	0.474 &  4.74E43 &    QSO     \\
  76   &     14:11:23.381 & 52:18:23.62 &  14:11:23.25 & 52:18:20.63 & 20.3  &  8.56E-16  &  1   &    n   &  	0.374 &  4.07E41 &    QSO     \\
  77   &     14:11:23.813 & 52:13:45.84 &  14:11:24.12 & 52:13:46.9  & 22.56 &  5.07E-16  &  1   &    n   &  	-     &	 - &	        \\
  79   &     14:11:24.503 & 52:19:23.85 &  14:11:24.75 & 52:19:23.99 & 20.83 &  6.56E-15  &  1   &    y   &  	-     &	 - &	        \\
  83   &     14:11:25.935 & 52:18:47.99 &  14:11:26.18 & 52:18:47.99 & 21.66 &  4.75E-15  &  1   &    y   &  	-     &	 - &	        \\
  87   &     14:11:28.499 & 52:19:29.85 &  14:11:28.36 & 52:19:30.08 & 22.45 &  1.88E-15  &  1   &    y   &  	-     &	 - &	        \\
  89   &     14:11:28.822 & 52:14:34.37 &  14:11:28.35 & 52:14:33.35 & 19.3  &  4.21E-16  &  1   &    n   &  	-     &	 - &	        \\
  91   &     14:11:29.495 & 52:16:39.51 &  14:11:29.7  & 52:16:40.22 & 22.42 &  4.27E-16  &  1   &    n   &  	-     &	 - &	        \\
  92   &     14:11:29.732 & 52:18:5.26  &  14:11:30.03 & 52:18:5.79  & 22.36 &  7.96E-16  &  1   &    y   &  	-     &	 - &            \\
  93   &     14:11:30.625 & 52:14:22.96 &  14:11:30.97 & 52:14:23.64 & 14.22 &  1.81E-15  &  1   &    n   &  	0     &  - &	Star    \\
  94   &     14:11:31.96  & 52:13:39.14 &  14:11:32.38 & 52:13:41.17 & 22.18 &  2.96E-15  &  1   &    n   &  	-     &	 - &            \\
  95   &     14:11:32.413 & 52:17:54.82 &  14:11:32.66 & 52:17:55.71 & 17.56 &  6.88E-16  &  1   &    n   &  	0.159 &  1.22E41 &	Gal     \\
  97   &     14:11:32.983 & 52:15:41.54 &  14:11:32.95 & 52:15:43.86 & 23.37 &  1.23E-15  &  1   &    n   &  	-     &	 - &	        \\
  98   &     14:11:33.121 & 52:12:56.34 &  14:11:33.48 & 52:12:57.07 & 23.0  &  4.43E-16  &  4   &    n   &  	-     &	 - &	        \\
 100   &     14:11:36.818 & 52:18:45.47 &  14:11:36.89 & 52:18:46.35 & 21.28 &  4.91E-15  &  1   &    n   &  	-     &	 - &            \\
 101   &     14:11:37.192 & 52:18:27.08 &  14:11:37.09 & 52:18:27.9  & 21.7  &  1.81E-15  &  1   &    y   &	-     &	-     &	     \\
 102   &     14:11:38.83  & 52:16:22.37 &  14:11:39.06 & 52:16:22.86 & 21.9  &  3.21E-15  &  1   &    y   &	-     &	-     &	     \\
 104   &     14:11:42.016 & 52:12:30.02 &  14:11:41.62 & 52:12:29.56 & 21.8  &  1.31E-15  &  4   &    y   &	-     &	-     &	     \\
 107   &     14:11:42.569 & 52:19:5.78  &  14:11:42.7  & 52:19:6.3   & 24.09 &  8.48E-16  &  1   &    y   &	-     &	-     &	     \\
 108   &     14:11:42.912 & 52:16:35.58 &  14:11:43.1  & 52:16:36.0  & 22.8  &  1.48E-15  &  1   &    n   &	-     &	-     &	     \\
 111   &     14:11:44.33  & 52:18:33.05 &  14:11:44.56 & 52:18:33.73 & 18.73 &  7.68E-16  &  1   &    n   &	-     &	-     &	     \\
 113   &     14:11:47.561 & 52:17:1.87  &  14:11:47.75 & 52:17:2.41  & 21.25 &  1.40E-15  &  1   &    n   &	-     &	-     &	     \\
 114   &     14:11:47.976 & 52:17:58.93 &  14:11:48.2  & 52:17:59.62 & 22.31 &  3.91E-15  &  1   &    y   &	-     &	-     &	     \\
 115   &     14:11:48.227 & 52:11:29.87 &  14:11:48.48 & 52:11:31.01 & 20.58 &  3.97E-15  &  4   &    n   &	-     &	-     &	     \\
 118   &     14:11:51.042 & 52:14:57.16 &  14:11:51.2  & 52:14:57.07 & 24.07 &  7.73E-16  &  1   &    y   &	-     &	-     &      \\
 122   &     14:11:55.052 & 52:18:22.68 &  14:11:54.73 & 52:18:24.25 & 22.31 &  1.41E-15  &  1   &    y   &	-     &	-     &	     \\
 123   &     14:11:55.038 & 52:11:34.12 &  14:11:55.3  & 52:11:35.46 & 21.7  &  2.06E-15  &  4   &    n   &	-     &	-     &	     \\
 124   &     14:11:55.399 & 52:18:2.77  &  14:11:55.56 & 52:18:3.48  & 20.28 &  7.22E-15  &  1   &    n   &	1.636 &	1.30E44     & QSO  \\
 126   &     14:11:56.267 & 52:16:37.02 &  14:11:56.3  & 52:16:36.7  & 23.44 &  7.81E-16  &  1   &    n   &	-     &	-     &	     \\
 128   &     14:11:57.648 & 52:14:10.31 &  14:11:57.77 & 52:14:11.59 & 21.68 &  1.16E-15  &  4   &    n   &	-     &	-     &	     \\
 131   &     14:12:0.065  & 52:13:29.57 &  14:11:59.54 & 52:13:29.35 & 20.08 &  5.10E-15  &  4   &    y   &     -     & -     &      \\
\hline
\end{tabular}

\normalsize
\end{table*}

\begin{figure}
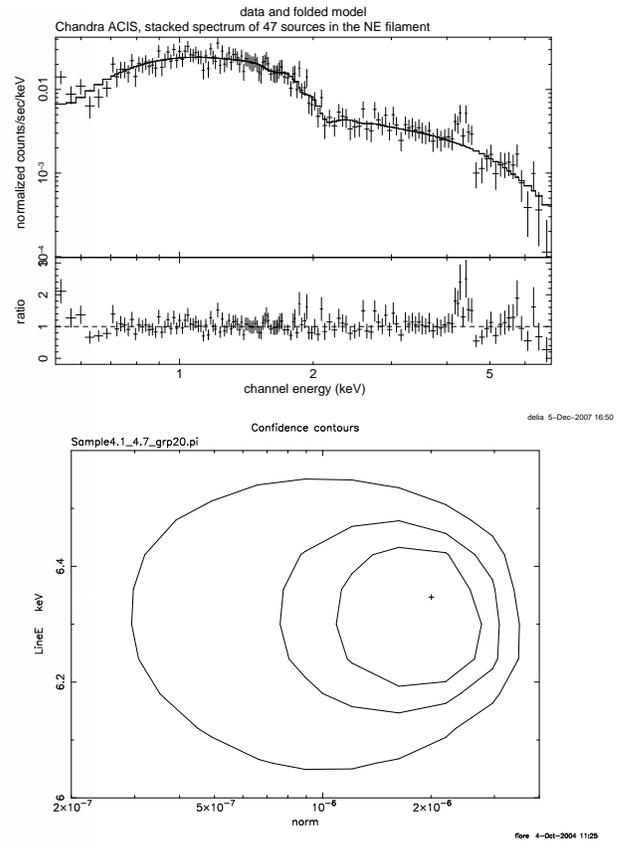

\centering
\includegraphics[angle=-90,width=8cm]{9338fig8.ps}
\includegraphics[angle=-90,width=7.5cm]{9338fig9.ps}
\caption{Upper panel: the stacked spectrum of the $47$ X-ray sources of Fig. 3
belonging to the 3C 295 field. A redshifted Fe line at E$\sim4.4$keV 
is clearly visible. Lower panel: the confidence contours of the line energy and
normalization; the rest frame energy of the line is in the range 
$6.2--6.47$ ($90\%$ confidence interval).}
\label{spe1}
\end{figure}


\section{X-ray Analysis}

We search for more hints to support the presence of a filament
spatially-connected to the 3C 295 cluster, using {\it Chandra}
data. Due to the sharpness of the {\it Chandra} Point Spread Function,
background counts at the position of the X-ray sources are often
negligible. Therefore, ``stacking'' together spectra extracted at the
position of each source almost increases linearly both the exposure
time and the sensitivity. This technique is particularly useful in our
case, because we are looking for a number of sources at similar
redshifts: individual spectral features would be significantly
enhanced in the stacked spectrum, if the sources are truly at the same
redshift. One of the strongest features in AGN X-ray spectra is the
iron k$\alpha$ line at 6.4 keV, which would be an obvious line to try
to identify in a stacked spectrum of {\it Chandra} sources.

\subsection{X-ray Data Preparation}

{\it Chandra} observed the $16'\times 16'$ field around the 3C 295
cluster using ACIS-I (Garmire 1997) on May 18, 2001. The aim point of
the observation was located at the position of 3C 295 $\alpha =$
14:11:10, $\delta=+$52:13:01 (J2000) and the exposure time was $\sim
92$ ks. D'Elia et al. (2005) identified $121$ sources in the $0.5-10$
keV band (see Fig. 3, top panel), using the PWDetect algorithm
(Damiani et al. 1997a, 1997b). We search here for the 6.4 keV iron
line emission at $z \sim 0.46$, and therefore extracted, from the
original data, images in the $4.1 \div 4.9 $ keV energy
range. Fifty-one of the 121 sources detected in the $0.5-10$ keV band
are detected in this image at a confidence level higher than $5
\sigma$. Four of these sources were spectroscopically-identified (see
Table 1) to be at a redshift different from that of 3C295 (see Fig. 3,
bottom panel). We excluded these sources from the analysis described
below. We were therefore left with $47$ sources.

\subsection{X-ray Stacked Spectrum}

We extracted a composite ``stacked'' spectrum from source region files
of radii $3$  arcsec centered at the positions of these $47$ sources. We
used the task {\em PSExtract} of the {\it Chandra} Interactive
Analysis of Observations (CIAO) software
(http://cxc.harvard.edu/ciao). A background spectrum was
extracted in several source-free regions distributed close to the $47$
sources.  Response matrices and effective-area files were created at
the centroid of the $47$ sources and added together to obtain average
responses.

The stacked spectrum was modeled using {\em XSPEC} version 11. We used
a model that included the following components: 1) a power law with
both energy index and normalization as free parameters; 2) Galactic
absorption toward the 3C 295 field with N$_H$ fixed at the value of
$1.33 \times 10^{20}$ cm$^{-2}$ (Dickey \& Lockman 1990); 3) a neutral
absorber at the redshift of 3C 295 ($z=0.46$) with column density free
to vary.  The fit of this model produces a reduced $\chi^2$ of $1.07$
for $142$ degrees of freedom.  The residuals, plotted in Fig. 4, show
a narrow (at the resolution of the Chandra ACIS-I) line at E$\sim4.4$
keV, the energy of the iron K$\alpha$ line at the redshift of the
cluster. A fit to the stacked spectrum using the same model as above
plus a Gaussian line at z=0.46 produces an improvement of the $\chi2$
value to $0.96$ for three additional parameters. This is significant
at the $99.8$ \% level, using the F test. The bottom panel of Fig. 4
shows the $\chi^2$ confidence contour for the normalization and energy
of this line. This plot confirms a detection significant at higher
than a 99.7\% confidence level and constrains the energy of the line to
the range $6.20 \div 6.47$ leV (90\% confidence interval), thus
excluding an identification with the $6.7$ keV, helium-like, iron line
from the hot intra-cluster medium at higher than a 4$\sigma$ confidence
level. The rest-frame equivalent width of the 6.4 keV K$\alpha$ line
is $EW=230 \pm 90$ eV.

We verified that the line in the stacked spectrum was not due to the
contribution of a single source. First, we excluded the type 1 AGN at
$z=0.474$ (Fig. 2) from the stacked spectrum . This did not
significantly change the quality of the fits and the results reported
above. Secondly, we verified that the $4.4$keV feature was not due to
a selection effect. For this purpose, we stacked the spectra of the 3
sources with an optically-identified redshift far from $z=0.46$ in the
energy range $ 4.1 \div 4.9 $ keV.  This spectrum does not show any
sign of a feature at $4.4$keV. The 90\% limit on the equivalent width
of a line at 4.4 keV is $200$ eV.  Thirdly, we divided the $ 4.1 \div
4.9$ keV sample into four groups, according to the location of the
sources in the four quadrants plotted in the bottom panel of
Fig. 3. We then performed a stacking analysis similar to that
performed for the complete sample. We find that the feature at $\sim
4.4$ keV is visible in the NE ($EW=150 \pm 100$ eV), NW ($EW=50 \pm
150$ eV), and SE ($EW=60 \pm 150$ eV) quadrants. The feature was not
detected, however, at a significance level higher than 99.7\% in any
of these spectra.  Although there is a hint that the iron K$\alpha$
equivalent width is higher for the NE quadrant, the quality of our
data is not sufficiently high to understand whether the signal
is mainly associated with sources in this quadrant or is distributed
more evenly in two or three quadrants. Deeper observations are clearly
required for this purpose.


\section{Conclusions and Discussion}

A $92$ ks {\it Chandra} observation of the field around the $z = 0.46
$ 3C 295 cluster of galaxy shows an excess of sources visible in the
NE quadrant (Fig. 3). D'Elia et al. (2005) performed a detailed study
of the field, applying three different methods of analysis, namely,
the logN-logS computed separately for the four ACIS-I chips, the
two-dimensional Kolmogorov-Smirnov (KS) test, and the 2 points angular
correlation function. All of these analysis methods suggest an asymmetric
distribution of sources in the NE region of the field and a strong
clustering on scales of a few arcmins.

Cappi et al. (2001) discussed four possible causes of these ``excess''
sources: (1) gravitationally-lensed very faint sources; (2) rapid
evolution of cluster AGNs or starburst galaxies (3) cosmic variance
of background sources; (4) LSS associated with the cluster.  Since
the surplus sources are not symmetrically distributed around the cluster,
our results rule out lensing by the cluster potential well and AGNs 
within the cluster.

In this paper, we studied in more detail the association of the excess sources
with the cluster. We used  two different approaches, namely 
a direct spectroscopic identification of the optical counterpart of the
X-ray sources, and X-ray stacked spectra.

Twenty-six out of $121$ X-ray sources were identified with an optical
counterpart brighter than $R=22$. Reliable redshifts were obtained for
$11$ sources.  Three objects were stars, two of which were located in
the SW quadrant of the {\it Chandra} observation, and the third was a
questionable identification. One type 1 AGN has a redshift very
similar to that of 3C295 and it is located in the NE quadrant, in
addition to two of the other three objects with redshifts in the range
$0.37 \div 0.53$. The remaining sources were two $z>1$ QSOs and two
low-redshift galaxies (z$<0.3$).  These results enhance the asymmetric
distribution found by D'Elia et al. (2005), since the two stars are
located in the SW, less-populated chip of the {\it Chandra} field.

Further evidence for an association of at least some of the ``excess''
sources with the LSS linked to 3C295 comes from the analysis of the
X-ray ``stacked'' spectrum of 51 {\it Chandra} sources detected in the
$4.1 \div 4.9 $ keV band. This band includes the redshifted iron
K$\alpha$ line at $z \sim 0.46$. We excluded objects with a
spectroscopic redshift significantly different from that of the
cluster.  We detected in the stacked spectrum an emission line at
$\sim4.4$ keV with a statistical significance better than 99.7 \%
confidence level.  Assuming the redshift equal to that of the cluster,
the line energy was measured to be in the range $6.20 \div 6.47$ keV
(90\% confidence interval). This range excludes the $6.7$ keV
helium-like iron line from the hot intracluser medium at better than 4
$\sigma$. We verified that the line in the stacked spectrum is not due
to the contribution of a single source. We conclude that the detection
of the redshifted line is a strong indication that at least several of
the excess sources lie at z$\sim0.46$, and have an AGN signature. They
are therefore associated with the central cluster and possibly form a
``filament'' connected with 3C 295.

At z=0.46, the 0.5-10 keV luminosity of these sources is in the range
$10^{42} \div 10^{44}$ ergs s$^{-1}$ (median $= 3 \times 10^{42}$ ergs
s$^{-1}$).  These luminosities are typical for Seyfert-like galaxies.
Martini et al. (2006), studying $40$ sources in the field of $8$ {\it
Chandra} low-redshift clusters ($0.05 < z < 0.31$), found that at
least $35$ of these objects have AGN signatures. Such signatures were
discovered not only using optical emission lines, but, above all, by
observing the X-ray spectral shapes and the multiwavelength flux
ratio. The authors concluded that dilution of low-luminosity AGN
spectral signatures by host galaxy starlight and obscuration of
accretion onto the central black hole did not easily allow these
sources to be optically classified as AGNs. As a consequence, the AGN
fraction in clusters of galaxies is a factor of $5$ higher than
previously observed. This agrees with our result, since the excess
sources close to 3C 295 have an AGN signature in X-rays, but most have
faint optical counterparts.

The virial radius, which is the radius at which the density contrast
is $2500$, of 3C 295 is $r=0.4$ Mpc (Allen, Schmidt \& Fabian). The
integration of the density profile up to this radius, yields a mass of
$2 \times 10^{14} M\odot$ and a velocity dispersion of $\sim 1300$ km
s$^{-1}$ (Dressler \& Gunn 1992). The allowed $\Delta$z compared to
these data is $\sim 0.005$. On the other hand, an association with 3C
295 defines a spatial scale of $2$ Mpc (5 arcmin) for the adopted
cosmology; this implies that we are sampling a cosmic filament of the
LSS and not cluster members.  Ruderman \& Ebeling (2005) analyzed the
spatial distribution of a sample of $508$ X-ray sources detected by
{\it Chandra} ACIS-I in 51 massive galaxy clusters ($z \sim 0.3 \div
0.7$). They found a significant ($8 \sigma$) excess within $3.5$ Mpc
from the cluster center, and they believed it to be caused by AGNs
connected to the central clusters.  They concluded that for relaxed
clusters, such as 3C 295, two spikes in the spatial distribution of
the X-ray sources emerged. The authors believed that the first scale
of $<0.5$ Mpc, was due to galaxy interactions involving the giant
ellipticals close to the central core, while the second scale at $2-3$
Mpc, was caused by galaxy mergers. These mergers are more likely to
occur in a transition region between the cluster and the field, where
the galaxy velocity dispersion is lower, but become rarer between the
two spatial scale peaks, where the velocities are higher. The
position of this second spike intriguingly agrees with our spatial
scale of $2$ Mpc; this can indicate that we are mapping a
transition region between the cluster and the field, that is, a cosmic
'filament', where pronounced AGN activity is present.

The integral of the field galaxy luminosity function at z=0.5
(e.g. Poli et al. 2001) provides $\sim 0.13$ galaxies Mpc$^{-3}$ for $M_B
< -17$, a resonable faint end optical luminosity, corresponding to our
lower X-ray luminosities. If we assume that roughly one tenth of the
galaxies are active X-ray sources of $L_X > 3 \times 10^{41}$ ergs
s$^{-1}$, then we would expect $\sim0.013$ X-ray sources Mpc$^{-3}$.
Since we calculate $\sim0.9$ sources Mpc$^{-3}$, this implies a galaxy
overdensity of $\approx70$, with of course a large (factor of 2-4)
positive and negative uncertainty, because of the uncertainties in our
space densities and assumptions. Again, this number is interestingly
close to the expected galaxy overdensity of filaments $\sim 10 \div
10^2$, and much smaller than the overdensities of clusters of galaxies
($\sim 10^3 \div 10^4$, the density contrast at the virial radius
being $\sim 2500$). This is further confirmation that we are probing a
cosmic 'filament' connected to the galaxy cluster 3C 295.

\begin{acknowledgements}
V.D. acknowledges support from ASI contract I/011/07/0, 
I/R/039/04 and I/R/023/05/0.
\end{acknowledgements}




\begin{thebibliography}{}

\bibitem[]{} Allen, S.W., Schmidt, R.W. \& Fabian, A.C., 2001, MNRAS, 328, L37

\bibitem[]{} Almaini, O., Scott, S.E., Dunlop, J.S. et al. 2003, MNRAS, 338, 303


\bibitem[]{}Best, P.N., van Dokkum, P.G., Franx, M \& Rottgering,
H.J.A., 2002, MNRAS, 330, 17

\bibitem[]{} Cappelluti, N., Cappi, M., Dadina, M., Malaguti, G., Branchesi, M., D'Elia, V. \& Palumbo, G.G.C., 2005, A\&A, 430, 39

\bibitem[]{}Cappi, M., Mazzotta, P., Elvis, M. et al. 2001, ApJ, 548, 624

\bibitem[]{}Daddi, E., Broadhurst, T., Zamorani, G., Cimatti, A., Roettinger, H.,
\& Renzini, A., 2001, A\&A  376, 825

\bibitem[]{}Damiani, F., Maggio, A., Micela, G. \& Sciortino, S.,
1997a, ApJ, 483, 350

\bibitem[]{}Damiani, F., Maggio, A., Micela, G. \& Sciortino, S.,
1997b, ApJ, 483, 370

\bibitem[]{}Dav\'e, R., Spergel, D.N., Steinhardt, P.J. \& Wandelt,
  B.D., 2001, ApJ, 547, 574

\bibitem[]{} D'Elia, V., Fiore, F, Elvis, M., Cappi, M., Mathur, S., Mazzotta, P., Falco, E., Cocchia, F., 2004, A\&A, 422, 11

\bibitem[]{}Dickey, J.M. \& Lockman, F.J., 1990, Ann. Rev. Ast. Astr.,ApJ...301..689H
28, 215.

\bibitem[]{}Dressler, A. \& Gunn, J.E., 1992, ApJS, 78, 1



\bibitem[]{}Fiore F., Nicastro, F., Savaglio, S., Stella, L. \&  Vietri, M.,
2000, ApJL, 544, 7 

\bibitem[]{}Garmire, G.P., 1997, AAS, 29, 283



\bibitem[]{}Giavalisco, M. \& Dickinson, M., 2001, ApJ, 550, 177 






\bibitem[]{}Martini, P., Kelson, D.D., Mulchaey, J.S. \& Trager, S.C.,
2002, ApJL, 576 109

\bibitem[]{}Martini, P., Kelson, D.D., Kim, E., Mulchaey, J.S. \& Athey, A.A., 2006, ApJ, 644, 116

\bibitem[]{}Mathur, S., Weinberg, D.H. \& Chen, X.
2003, ApJ, 582, 82

\bibitem[]{}Molnar, S.M., Hughes, J.P, Donahue, M. \& Joy, M., 2002,
ApJL, 573, 91

\bibitem[]{}Nicastro, F., Zezas, A., Drake, J. et al., 2002, ApJ, 573, 157

\bibitem[]{}Nicastro, F., Zezas, A.,  Elvis, M. et al., 2003, Nature,
  421, 719


\bibitem[]{}Peacock, J.A. 1999, ``Cosmological physics'' (Cambridge:
Cambridge University Press)


\bibitem[]{}Pentericci, L., Kurk, J.D., Carilli, C.L., Harris, D.E.,
Miley, G.K. \& Rottgering, H.J.A., 2002, astro-ph/0209392

\bibitem[]{}Poli, F.,  Giallongo, E., Fontana, A.,
  Cristiani, S. \& D'Odorico, S., 2001, ApJL, 551, 45

\bibitem[]{} Ruderman, J.T. \& Ebeling, H., 2005, ApJ, 623, 81L 



\bibitem[]{}Soltan, A.M., Freyberg, M.J. \& Hasinger, G., 2002, A\&A, 395, 475




\bibitem[]{}Yang, Y., Mushotzky, R.F., Barger, A.J., Cowie, L.L,
  Sanders, D.B., \& Steffen, A.T., 2003, ApJL, 585, 85


\bibitem[]{}Zappacosta, L., Mannucci, F, Maiolino, R. et al., 2002, A\&A, 394, 7

\end{thebibliography}
\end{document}